\documentclass[nonacm]{acmart}
\usepackage{amsmath,amsfonts}
\usepackage{graphicx}
\usepackage{textcomp}
\usepackage{xcolor}
\usepackage{amsmath}
\usepackage{listings}
\usepackage{algorithm}
\usepackage[noend]{algpseudocode}
\usepackage{subcaption}
\usepackage[normalem]{ulem}
\usepackage{csquotes}
\usepackage{tikz}
\def\BibTeX{{\rm B\kern-.05em{\sc i\kern-.025em b}\kern-.08em
    T\kern-.1667em\lower.7ex\hbox{E}\kern-.125emX}}
    
\newcommand{\SystemName}{ASTANA}

\newcommand*\circled[1]{\tikz[baseline=(char.base)]{
            \node[shape=circle,draw,inner sep=2pt] (char) {#1};}}

\definecolor{codegreen}{rgb}{0,0.6,0}
\definecolor{codegray}{rgb}{0.5,0.5,0.5}
\definecolor{codepurple}{rgb}{0.58,0,0.82}
\definecolor{backcolour}{rgb}{0.98, 0.98, 0.97}

\lstdefinestyle{java}{
    backgroundcolor=\color{backcolour},   
    commentstyle=\color{codegreen},
    keywordstyle=\color{magenta},
    numberstyle=\tiny\color{codegray},
    stringstyle=\color{codepurple},
    basicstyle=\ttfamily\footnotesize,
    breakatwhitespace=false,         
    breaklines=true,                 
    captionpos=b,                    
    keepspaces=true,                 
    numbers=left,                    
    numbersep=5pt,                  
    showspaces=false,                
    showtabs=false,                  
    tabsize=2,
    escapechar=\$,
    frame=single,
}

\lstdefinestyle{smali}{
    backgroundcolor=\color{backcolour},   
    commentstyle=\color{codegreen},
    keywordstyle=\color{magenta},
    numberstyle=\tiny\color{codegray},
    stringstyle=\color{codepurple},
    basicstyle=\ttfamily\footnotesize,
    breakatwhitespace=false,         
    breaklines=true,                 
    captionpos=b,                    
    keepspaces=true,                 
    numbers=left,                    
    numbersep=5pt,                  
    showspaces=false,                
    showtabs=false,                  
    tabsize=2,
    escapechar=\$,
    float,
    floatplacement=tbp,
    frame=single,
}

\lstdefinelanguage{smali}{
    alsoletter=-/,
    morekeywords={const/16,const-string,invoke-static,move-result,xor-int/2addr,if-eqz,add-int/2addr,int-to-short},
    sensitive=false, 
    morecomment=[l]{//}, 
    morecomment=[s]{/*}{*/}, 
    morestring=[b]" 
} %
 
 \makeatletter
 \renewcommand{\ALG@beginalgorithmic}{\small}
 \makeatother
 
\begin{document}

\title{\SystemName: Practical String Deobfuscation for Android Applications Using Program Slicing}

\author{Martijn de Vos and Johan Pouwelse}
\affiliation{\institution{Delft University of Technology}}


\begin{abstract}
	Software obfuscation is widely used by Android developers to protect the source code of their applications against adversarial reverse-engineering efforts.
	A specific type of obfuscation, \emph{string obfuscation}, transforms the content of all string literals in the source code to non-interpretable text, and inserts logic to deobfuscate these string literals at runtime.
	
	In this work, we demonstrate that string obfuscation is easily revertible.
	We present \SystemName, a practical tool for Android applications to recovers the human-readable content from obfuscated string literals.
	\SystemName~makes minimal assumptions about the obfuscation logic or application structure.
	The key idea is to execute the deobfuscation logic for a specific (obfuscated) string literal, which yields the original string value.  
	To obtain the relevant deobfuscation logic, we present a lightweight and optimistic algorithm, based on program slicing techniques.
	
	By an experimental evaluation with 100 popular real-world financial applications, we demonstrate the practicality of \SystemName.
	We verify the correctness of our deobfuscation tool and provide insights in the behaviour of string obfuscators applied by the developers of the evaluated Android applications.
\end{abstract}

\maketitle

\section{Introduction}
Malign efforts to reverse engineer deployed mobile applications are a serious threat to both companies and mobile users~\cite{gonzalez2015exploring}.
On one hand, malware authors commonly plagiarize benign Android applications by reverse engineering of their application logic, in order to inject malicious code.
The modified application is then redeployed, often under a similar name.
On the other hand, companies are motivated to reverse engineer competitive applications to gain access to intellectual property like copyrighted assets or proprietary algorithms~\cite{Wermke2018ALS}.

To secure mobile applications against adversarial reverse engineering, developers often \emph{obfuscate} the source code of their applications prior to deployment.
Obfuscation is a software transformation that aims to protect some deployed application by complicating or misleading an analysis of their content. 
Commonly used obfuscation techniques include identifier renaming (e.g., replacing meaningful identifiers with meaningless information), control flow modification (e.g., by inserting useless conditionals, making the program flow harder to comprehend), and string obfuscation.

\begin{lstlisting}[style=java,language=Java,float=tp,caption={An example of a common string obfuscation. The string obfuscator encrypts declared string literals and inserts code to decrypt this string during runtime.}, label=lst:stringobfuscation]
	public void someMethod(int a, String b) {
		$\sout{String secret = "MY\_SECRET\_KEY";}$
		String secret = "YkfV^SSQ?P]\Q";
		int decryptionKey = 58934;
		secret = Decryptor.decrypt(secret, decryptionKey);
		...
	}
\end{lstlisting}

\emph{String obfuscation} is a technique to masquerade human-readable string literals that have been defined in the source code.
These strings may leak sensitive information about the application, e.g., the usage of cryptographic primitives, encryption keys, or even personal information of developers~\cite{zhou2015harvesting}.
Although string obfuscation is not widely applied for benign applications, is often used by malware authors to avoid detection of their malign applications by automated malware scanners.
These scanners inspect string literals to indicate the presence of malicious code, amongst other checks.
An example of a common approach for string obfuscation is given in the Java code in Listing~\ref{lst:stringobfuscation}.
On line 2, the programmer has defined a string literal with content \enquote{\texttt{MY\_SECRET\_KEY}}, stored in a variable named \texttt{secret}.
An obfuscation tool transforming this literal has first encrypted the string literal stored in the \texttt{secret} variable, using an unknown encryption scheme, and then inserted additional logic (line 3, 4 and 5) to decrypt it on runtime.
Decryption is carried out by the static \texttt{decrypt} method on line 5. 


Although various commercial obfuscators provide string obfuscation, we argue that their transformation is not effective.
The reason is that most string obfuscators insert similar logic when obfuscation string literals within an application, making it trivial to detect the usage of string obfuscation. 
While there exists research on detecting string obfuscation, there are very few studies or functional tools that revert the original content of obfuscated string literals.
We observe that the few existing approaches have low context-awareness and often fail when other obfuscation techniques have been applied, in particular control flow obfuscation.
The aim of this work is to devise a generic approach to deobfuscate string literals in Android applications while making minimal assumptions about the used obfuscator and application logic.

We present \SystemName, a string deobfuscation tool for Android applications.
\SystemName~extracts the deobfuscation logic for each string literal and executes this logic to retrieve the original string value in non-obfuscated form.
The logic is extracted using a lightweight, optimistic program slicing algorithm.
\SystemName~functions even when the control flow is heavily modified, in contrast to existing string deobfuscation tools.
An experimental evaluation with 100 real-world financial applications demonstrates the practicality of \SystemName.
The contributions of this work can be summarized as follows:


\begin{itemize}
\item \SystemName: A string deobfuscation tool for Android applications using program slicing (Section~\ref{sec:approach}).
\item A functional, open-source implementation of \SystemName, in the Java programming language (Section~\ref{sec:implementation_evaluation}).
\item An experimental evaluation, showing the practicality of our tool on real-world Android applications (Section~\ref{sec:implementation_evaluation}).
\end{itemize}

\section{Background and Problem Description}
The aim of this work is to reverse the obfuscation of string literals in Android applications using a generic approach.
Our high-level approach consists of three phases: first, we identify and mark string literals which are likely to be obfuscated.
Second, we extract the statements that are involved in the deobfuscation of a specific string literal.
Finally, these snippets are executed to retrieve the original string literal.
We use \emph{backwards program slicing} to extract the logic that deobfuscates a specific string literal.
This concept is briefly elaborated next.

\subsection{Backwards Program Slicing}
Backwards program slicing is a technique in software engineering to compute a subset of statements, or slices, of a particular method, with respect to a statement $ s $ and a variable of interest $ V $.
This combination is also called the \emph{slicing criterion} and denoted in this work as the two-tuple $ C = (s, V) $.
All data flows in a backwards slice end in the slicing criterion.
In other words, a backwards slice contain all statements that eventually \enquote{influence} the slicing criterion.
Common applications include the detection of code duplications and the removal of dead (unreachable) code.
To deobfuscate string literals, we propose to use the statement where a string literal is decrypted as $ s $, and the variable that stores the non-obfuscated string as our variable of interest, $ V $.
Referring to the code in Listing~\ref{lst:stringobfuscation}, our slicing criterion would be $ C = (5, secret) $.
A backwards slice based on $ C $ would then include the statements on line 3, 4 and 5.

An important distinction is made between \emph{static} and \emph{dynamic} program slicing.
Static slicing does not consider a specific input for the program being sliced, while a dynamic slicing algorithm performs slicing under a specific input (or a specific execution).
In this work, we specifically opt for static slicing since it is infeasible to obtain numerous execution traces for a given application in a fully automated manner.
For example, banking applications require user credentials that are not always easy to obtain, thus complicating this process.

\subsection{Problem Description}
\label{sec:problem_description}
We identify two main problems when applying backwards program slicing to deobfuscate string literals.

\textbf{Problem 1: Accurately determine slicing criteria.}
Generating a backwards program slice requires us to define a slicing criterion $ (s, V) $.
As discussed, the slicing criterion to generate this program slice would ideally be located at the statement right after deobfuscation of the string literal of interest.
Based on the location of the deobfuscation logic, the slicing criterion can be established.
The main problem, however, is to identify where a specific string literal is possibly deobfuscated. 
\SystemName~requires a mechanism to determine potential \emph{deobfuscation candidates} in order to determine slicing criteria and to construct program slices.

\textbf{Problem 2: Slicing complex methods.}
Many existing algorithms derive (static) backwards program slices from the Program Dependence Graph (PDG)~\cite{Ottenstein1984ThePD}.
The PDG is a directed graph for a single method where each node corresponds to a statement, and a directed edge between two nodes indicate either a data or a control dependency.
The PDG is a combination between a data dependency graph and a control flow graph.
Determining a backwards program slice for a slicing criterion $ (s, V) $ can be achieved by including the statements for all nodes in the PDG that are reachable from the node corresponding to $ s $.

Although the PDG is a reliable data structure to generate program slices, the computational complexity required to build the data structure for individual methods is a major limitation~\cite{Mohammadian2018IntraproceduralSU}.
Specifically, the computational complexity to derive a PDG grows exponentially with the number of control flow elements in a method.
Furthermore, since the number of methods in Android applications can be large, we consider the PDG not suitable for practical string deobfuscation.
Therefore, we aim for a more efficient approach to determine program slices, specifically when the considered method is complex.

\begin{figure*}[t]
	\centering
	\includegraphics[width=.9\linewidth]{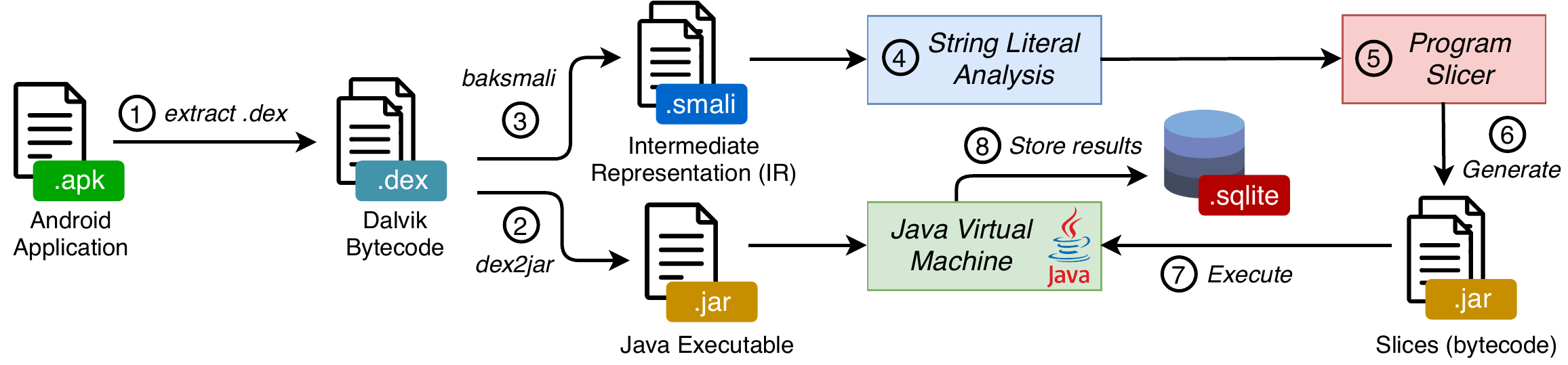}
	\caption{An high-level overview of \SystemName, our tool to deobfuscate string literals in Android applications using program slicing.}
	\label{fig:deobfuscation_process}
\end{figure*}

\section{Practical String Deobfuscation}
\label{sec:approach}
We now describe our string deobfuscation approach, and show how \SystemName~overcome the problems posed in Section~\ref{sec:problem_description}.
Figure~\ref{fig:deobfuscation_process} shows a high-level overview of \SystemName.
The visualized steps are explained next.






\subsection{Parsing the APK file}
To deobfuscate string literals in a specific Android application, \SystemName~expects an APK file as input.
The APK file, or Android Package Kit, contains (compiled) Dalvik-compatible bytecode, executed by the Dalvik Virtual Machine when the user starts an application.
Furthermore, the APK contains auxiliary resource files such as images or audio files, which are ignored by \SystemName.
\SystemName~first extracts the bytecode using the open-source \emph{dex2jar} library\footnote{https://github.com/pxb1988/dex2jar}, see step \circled{1} in Figure~\ref{fig:deobfuscation_process}.
This results in a collection of \emph{.dex} files.

\begin{lstlisting}[language=smali,style=smali,caption={A snippet of Smali code, showing a subset of implementation of a method named \emph{Bjg}. An obfuscated string literal is defined on line 5 and deobfuscated (decrypted) on line 17.}, label=lst:smaliexample]
.method public final Bjg()V
    ...
    const/16 v1, 31930
    const/16 v2, 3479
    const-string v3, "=ZfZ[a"
    invoke-static { }, Lu/VnN;->T()I
    move-result v0
    xor-int/2addr v0, v1
    if-eqz v0 :L1
    add-int/2addr v1, v0
  :L1
    int-to-short v1, v0
    invoke-static { }, Lu/VnN;->T()I
    move-result v0
    xor-int/2addr v0, v2
    int-to-short v0, v0
    invoke-static { v3, v1, v0 }, Lu/NS;->b(Ljava/lang/String;SS)Ljava/lang/String;
    move-result-object v0
    ...
\end{lstlisting}

The resulting \emph{.dex} files are then used to generate two intermediate artifacts.
First, \SystemName~converts the Dalvik bytecode to Java-compatible bytecode and saves this code in a single JAR file, see step \circled{2}.
This JAR file is used by the Java Virtual Machine when executing program slices to obtain the deobfuscated string literals, see Section \ref{sec:slice_execution}.
The conversion from Dalvik bytecode to Java bytecode enables \SystemName~to conduct the deobfuscation process on non-mobile devices, since the execution of Dalvik bytecode is usually limited to devices running the Android OS.
Running the deobfuscation in the Java Virtual Machine is viable since deobfuscation logic often does not invoke Android-specific logic.

Besides the generation of Java-compatible bytecode, \SystemName~also convert the \emph{.dex} files to a collection of files containing Smali.
This conversion is performed by the \emph{baksmali} tool, a component of the \emph{dex2jar} library to convert Dalvik bytecode to Smali.
Smali is a human-readable intermediate representation of Dalvik bytecode and is often used during Android application analysis~\cite{Wang2015WuKongAS}.
This is exemplified in Listing~\ref{lst:smaliexample}, which shows a subset of all statements in the \texttt{Bjg} method (its method signature is located on line 1).
In smali, a single statement corresponds to one of the 255 available Dalvik opcodes.
Listing~\ref{lst:smaliexample} highlights the human-readable operation names corresponding to the opcodes, e.g., \texttt{invoke-static} to invoke a static method, or \texttt{int-to-short} to convert an integer type to a short type.
Literals and object references are stored in registers (Listing~\ref{lst:smaliexample} uses registers \texttt{v0} to \texttt{v3}).
The Java code corresponding to the Smali snippet is given in Listing~\ref{lst:javaexample}.
In Listing~\ref{lst:javaexample}, registers have been replaced by variables, and variables have been given a more descriptive name.

The statements in Listing~\ref{lst:smaliexample} define an obfuscated string literal on line 5 which is passed to a static method \texttt{b} on line 17.
The return value of this method invocation is stored in register \texttt{v0} on line 18.
Note that this return value is likely to contain the string literal in non-obfuscated format, since the method \texttt{b} expects the obfuscated string literal as parameter, and the return type of the static method \texttt{b} is a \texttt{String} object.
The static method \texttt{b} is invoked with two other parameters: \texttt{v1} and \texttt{v0}.
These parameters have the primitive \texttt{short} type and can be interpreted as numeric decryption keys.
The Smali code in Listing~\ref{lst:smaliexample} will be a guiding example to illustrate the next steps in the process, where we determine deobfuscation candidates for string literals and perform backwards program slicing.

\begin{lstlisting}[style=java,float,float,floatplacement=tbp,language=Java, caption={The Java code, corresponding to the Smali snippet in Listing~\ref{lst:smaliexample}.}, label=lst:javaexample]
public final void Bjg() {
  ...
  int decKey1 = 31930;
  int decKey2 = 3479;
  String encrypted = "=ZfZ[a";
  int i = VnN.T();
  i ^= decKey1;
  if(i == 0) {
    decKey1 += i;
  }
  short i2 = (short) i;
  int i3 = VnN.T();
  i3 ^= decKey2;
  String decrypted = NS.b(encrypted, i2, i3);
  ...
}
\end{lstlisting}

\subsection{String Literal Analysis}
Given a collection of Smali files, \SystemName~now analyses string literals and attempts to find deobfuscation candidates, see step \circled{4} in Figure~\ref{fig:deobfuscation_process}.
Each Smali file hosts the implementation of exactly one (inner) class and its methods.
\SystemName~first iterates through all available methods in these classes.
We explicitly exclude popular open-source libraries and frameworks from the analysis, such as \emph{okhttp} and \emph{volley}.
Since the source code of these dependencies is publicly available, we consider the deobfuscation of their string literals not very useful.
The deobfuscation algorithm for the string literal analysis and the program slicing is provided in Algorithm~\ref{alg:stringdeobfuscation}.

Whenever \SystemName~analyse a method $ m $, it first devises the control-flow graph (CFG) of $ m $, referred to as $ CFG_m $.
The CFG is a directed graph that captures all possible program flows within a particular method.
Each node in the CFG represents a statement, and each edge $ (a, b) $ between two nodes (statements) indicates that the program flow can proceed from statement $ a $ to $ b $.
For example, the node corresponding to the conditional statement on line 9 in Listing~\ref{lst:smaliexample} contains two outgoing edges: one that redirects the program flow to line 12 (when the value in \texttt{v0} is zero), and another one that redirects the program flow to line 10 (when the value in \texttt{v0} is non-zero).
Constructing the CFG for a particular method induces time complexity $ O(n) $, with $ n $ being the number of statements in the method.

Next, \SystemName~performs a linear scan of all statement within the methods and identifies all string literals defined with the \texttt{const-string} or \texttt{const-string/jumbo} operation\footnote{this opcode is used when over 65.536 ($ 2^{32} $) strings have been defined in the Android application.} opcodes.
Declarations of empty strings (with length zero) are ignored since they are not likely to be obfuscated.
The set of all string literals defined in method $ m $ is denoted by $ L_{s,m} $.

By conducting a forward breadth-first search in $ CFG_m $ for each string literal $ s \in L_{s,m} $, \SystemName~locates all statements that could deobfuscate the string literal, also called \emph{deobfuscation candidates}.
This search starts at the statement immediately after $ s $ and terminates when the end of the method has been reached.
Since the search considers all possible flow redirections for conditional statements in $ m $, this search can take a long time if the method contains many conditionals.
Therefore, we limit the search scope such that we consider at most five conditional statements.
We believe this is reasonable, since inserted deobfuscation logic is not likely to be complex.
The resulting deobfuscation candidates for a string literal $ s $ is defined as $ D_{s} $.
We include a statement $ t $ in $ D_{s} $, if it meets one of the following three conditions:

\begin{enumerate}
\item $ t $ initializes a new \texttt{String} object by calling its \texttt{init} (constructor) method.
\item $ t $ is an invocation on a static method that returns a \texttt{String} object.
\item $ t $ attempts to casts an object of type \texttt{Object} to \texttt{String} using the \texttt{check-cast} opcode.
\end{enumerate}

Note how the three conditions above are likely to include statements that are \enquote{transforming} the string literal $ s $.
Although we employ a rule-based approach to devise the set $ D_s $, our evaluation in Section~\ref{sec:implementation_evaluation} shows that the three conditionals mentioned above are sufficient to detect the deobfuscation logic of most obfuscated string literals.
The reason is that string obfuscation tools only use a limited number of different string encryption mechanisms: our manual inspection only revealed around five different string obfuscation methods.

\subsection{Program Slicer}
\SystemName~now generates a backwards program slice for each deobfuscation candidates in $ D_{s} $, see step \circled{5} in Figure~\ref{fig:deobfuscation_process}.
First, we elaborate how the slicing criterion is determined.
Next, we discuss our approach to perform backwards slicing.

\textbf{Slicing Criterion. }
For each decryption candidate $ t \in D_{s} $, we determine the slicing criterion $ C_t = (i, V) $ as follows.
If $ t $ meets condition 1, we fix $ i $ to the first statement after $ t $, and fix our variable of interest, $ V $, to the first parameter of the call to the \texttt{String} constructor.
Note that it is likely that $ V $ contains a reference to the string literal $ s $ in non-obfuscated form, if string obfuscation is used.
If $ t $ meets condition 2, we use the corresponding \texttt{move-result-object} statement for $ i $, and the register where the resulting object is moved to as $ V $.
Since the statement on line 17 in Listing~\ref{lst:smaliexample} meets condition 2, the slicing criterion would be $ C = (18, v0) $.
Finally, if $ t $ meets condition 3, we use $ i = t $ and fix $ V $ to the register that stores the cast object.

\textbf{Backwards Slicing. }
Based on a deobfuscation candidate $ t \in D_{s} $, and a corresponding slicing criterion $ C_t = (i, v) $, \SystemName~now determines a backwards program slice.
In Section~\ref{sec:problem_description}, we motivated our decision to not rely on the Program Dependency Graph to determine these slices, since they are computationally intensive to compute for methods with a high cyclomatic complexity.

Instead, we devise a lightweight, optimistic approach to determine program slices.
The main approach of our algorithm is to first analyse the data flows from the deobfuscation candidates, to the statement with the string literal $ s $.
We believe that it is likely that the statements on this path are sufficient to generate a program slice that is able to deobfuscate a given string literal.
Only if the resulting program slice lacks required information, we extend the search scope and attempt to retrieve the missing information in the rest of the method logic.

We now describe our algorithm in more detail, and show how we determine program slices.
\SystemName~first determines all execution paths from statement $ s $ to $ t $ in method $ m $ by conducting a search in $ CFG_m $.
The set of all execution paths from statement $ s $ to $ t $ is denoted by $ P_{s, t} $.
Note that $ P_{s, t} $ contains at least one element: if $ P_{s, t} = \emptyset $, it would not have found the decryption candidate $ t $ in the first place.

We build a \emph{register dependency graph} (RDG) for each execution path $ p \in P_{s, t} $ and denote this graph by $ G_{reg,p} $.
The RDG is a directed graph which captures temporal dependencies between the state of registers used when executing the statements in an execution path. 
Each node is labelled by a two-tuple $ (n, i) $ where $ n $ is the register name, and $ i \in \mathbb{Z} $ is a counter that is incremented by one every time the register $ n $ is overwritten with a new value.
A directed edge from node $ a $ to node $ b $ denotes that the register value described by node $ b $ is dependent on the register value described by node $ b $.
$ G_{reg,p} $ is iteratively constructed by analysis of the statements in the execution path $ p $, starting from statement $ s $ and ending with statement $ t $.
During this construction, new nodes and edges are inserted accordingly, based on the logic of each analysed statement.
We have carefully analysed how each of the 255 available Dalvik opcode\footnote{Also see https://source.android.com/devices/tech/dalvik/dalvik-bytecode} results in new register state dependencies.
For example, the \texttt{add-int/2addr} opcode on line 10 in Listing~\ref{lst:smaliexample} overrides the register value of \texttt{v0} with the result of the addition of the current register values in \texttt{v0} and \texttt{v1}.
In the RDG, this opcode creates a dependency from the new register value of \texttt{v0} to the old register value of \texttt{v0} and the register value of \texttt{v1}.
While building the RDG, \SystemName~builds a mapping from each node in the RDG to a set of statements that use the register state captured by the node.
As we will show, the RDG is a key data structure when computing program slices.

\begin{figure}[t]
	\centering
	\begin{subfigure}[t]{.5\columnwidth}
		\centering
		\captionsetup{width=.9\columnwidth}
		\includegraphics[width=.7\columnwidth]{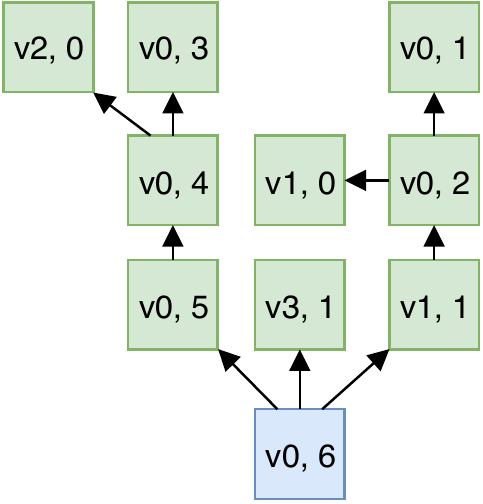}
		\caption{Without jump.}
		\label{fig:register_dependency_graph_with}
	\end{subfigure}%
	\begin{subfigure}[t]{.5\columnwidth}
		\centering
		\captionsetup{width=.9\columnwidth}
		\includegraphics[width=.7\columnwidth]{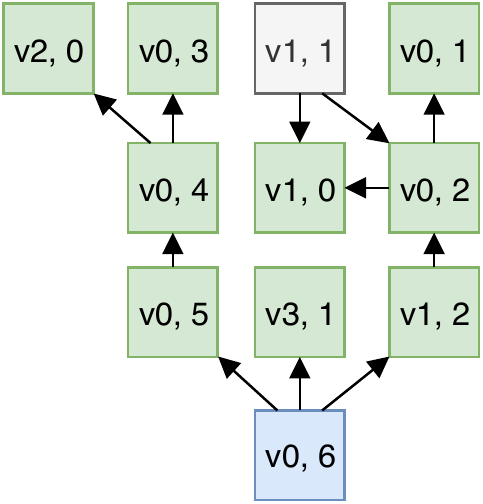}
		\caption{With jump.}
		\label{fig:register_dependency_graph_without}
	\end{subfigure}
	\caption{The register dependency graphs for the Smali code in Listing x (line 5 to line 18). The blue nodes correspond to line 18 and the green nodes form all registers that are reachable from the blue node.}
	\label{fig:register_dependency_graph}
\end{figure}

Figure~\ref{fig:register_dependency_graph} shows the register dependency graph for the two execution paths from line 5 up to and including line 18 in Listing~\ref{lst:smaliexample}.
Figure~\ref{fig:register_dependency_graph_without} shows the RDG for the execution path where the conditional at line 9 evaluates to false, whereas in the RDG of Figure~\ref{fig:register_dependency_graph_with} this conditional evaluates to true.
Note how the content of register \texttt{v1} is overwritten twice when the conditional on line 9 evaluates to true: on line 10 and line 11.
The node corresponding to the slicing criterion $ C_t = (18, v0) $ is coloured blue.
Note that when the program flow reaches the statement associated with the slicing criterion, the register value of \texttt{v0} has been overwritten six times.

After all RDGs have been computed for each $ p \in P_{s, t} $, \SystemName~verifies with a breadth-first search in $ p $ whether there is a path in all RDGs from the node associated with our slicing criterion $ C_t $ to the register that represents the string declaration $ s $.
For all RDGs, there should be a path from our slicing criterion to the statement that declares the string literal.
If no such path exists in at least one of the RDGs for $ P_{s, t} $, our suspected deobfuscation candidate does not rely on the string literal $ s $, directly nor indirectly.
In other words, deobfuscation candidate $ t $ does not require the information in the string literal $ s $.
In this situation, no program slice will be computed for deobfuscation candidate $ t $ and the process is repeated with another deobfuscation candidate in $ D_s $, if one exists.

Based on the computed RDGs for each $ p \in P{s, t} $, \SystemName~determines a program slice for a given slicing criterion $ C_t $ as follows.
For each PDG $ G_{reg} $, we mark all nodes that are reachable from the node corresponding to the slicing criterion.
For each marked node, we include all statements that correspond to the register value tracked by this node in the program slice (recall that this mapping is stored during the construction of a PDG).
Furthermore, we include all conditional opcodes that are encountered in any execution path in the program slice, such as if-statements, \texttt{goto} statements, and switch statements.

Although our approach generates a program slice with statements between $ s $ and $ t $ that are likely to be relevant for deobfuscation, we identify a limitation.
If we generate a program slice using the approach described above, the program slice taken from Listing~\ref{lst:smaliexample} does not include the constant definitions on line 3 and 4, since these statements are not between $ s $ and $ t $ in the program flow.
The resulting program slice is invalid since the string deobfuscation method \texttt{b} requires the constant declarations on line 3 and 4.
To address this situation, we enhance the PDG construction and keep track of registers that are not initialized in the program slice.
For instance, when generating the PDGs shown in Figure~\ref{fig:register_dependency_graph}, the initial register values of \texttt{v1} and \texttt{v2} are marked as undefined.
In other words, if there are any undefined variables, we extend the existing program slice corresponding to slicing criterion $ C_t $ with another slice, computed from the beginning of the method to statement $ s $.

Enhancing the program slice with this additional information results in another limitation: if there are many different paths from statement $ s $ to the beginning of the method, the process of determining execution paths might not terminate within reasonable time.
However, we note that not \emph{all} execution paths might be required to obtain an accurate program slice with all required statements.
Specifically, the differences in program flow between different execution paths is likely to be small.
If the queue of the breadth-first search from the beginning of the method to statement $ s $ exceeds a certain size, we change the search strategy to depth-first.
Since we already know that the number of possible execution paths is likely to be large, we terminate the depth-first search if a certain number of paths have been found.

After \SystemName~has determined the involved statements for a program slice, the program slice is written away to a Java class file, see Step \circled{6} in Figure~\ref{fig:deobfuscation_process}.
\SystemName~adds statements at the end of the program slice to write the deobfuscated string to the standard output, using the \texttt{System.out.println} method.
Since some methods that deobfuscate a given string verify the call stack on runtime, the program slice is embedded in the same method and class as the string literal that \SystemName~is trying to deobfuscate.

\subsection{Slice Execution}
\label{sec:slice_execution}
Finally, the generated Java class file is executed (step \circled{7}) in the Java Virtual Machine.
The execution environment has access to the class files contained in the JAR file generated during step \circled{2}.
These files are of paramount importance for our deobfuscation process, since they contain the implementation of deobfuscation methods.
We limit the execution time of the deobfuscation logic to five seconds, which we believe is a generous time window to deobfuscate a single string literal.
If the Java process exists with a non-zero exit code while trying to deobfuscate string literal $ s $, with deobfuscation candidates $ D_{s} $, we repeat the process and generate a new program slice with another deobfuscation candidate.
All results, including standard output, error messages, and the exit code of the process are committed to a local sqlite database, see step \circled{8}.

\section{Implementation and Evaluation}
\label{sec:implementation_evaluation}
We have implemented the \SystemName~string deobfuscation tool in the Java 8 programming language.
Our implementation spans 1.778 lines of code.
All source code, documentation and unit tests are published in a GitHub repository\footnote{See \url{https://github.com/devos50/astana}}.
As discussed in Section~\ref{sec:approach}, \SystemName~uses the \emph{dex2jar} library for the APK analysis, see steps \circled{1} to \circled{3} in Figure~\ref{fig:deobfuscation_process}.

\subsection{Evaluation}
We evaluate the implementation of~\SystemName~and show its performance on real-world financial Android applications.

\textbf{Dataset.}
We download the top 100 financial Android applications in the United Kingdom from the Google Play Store, based on statistics by AppAnnie\footnote{See \url{https://www.appannie.com/en/apps/google-play/top/}}.
Our evaluation specifically focusses on financial applications since they pose a valuable target for malware authors and therefore are likely to use obfuscation techniques.
We remark that our evaluation lacks ground truth regarding whether the applications under consideration has applied obfuscation of string literals.

\begin{figure}[t]
	\centering
	\includegraphics[width=.7\linewidth]{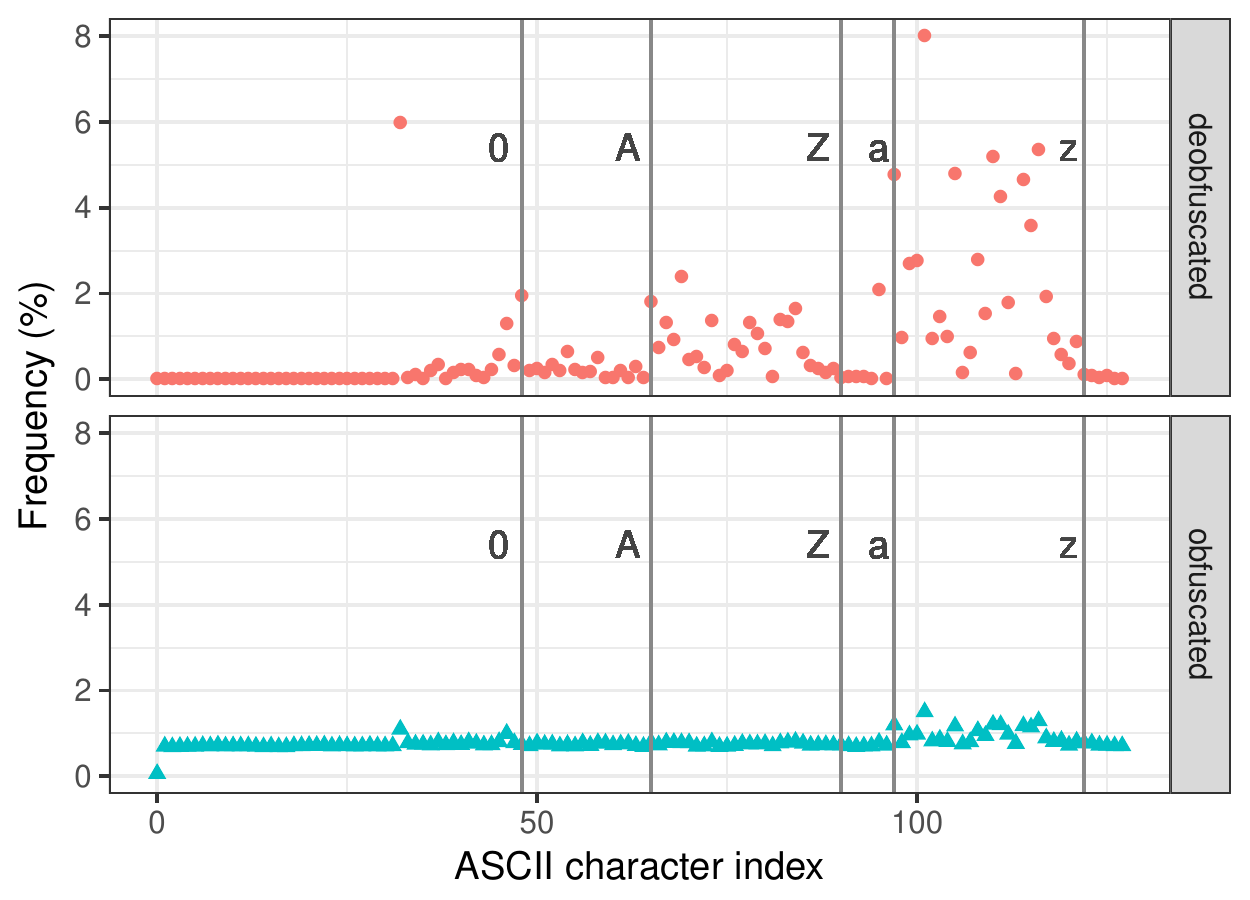}
	\caption{The distribution of ASCII characters in strings before deobfuscation (top plot) and after deobfuscation (bottom plot).}
	\label{fig:character_distribution}
\end{figure}

\textbf{Deobfuscation results.}
We use \SystemName~to analyse the collected financial Android applications, and attempt to deobfuscate string literals using the process visualized in Figure~\ref{fig:deobfuscation_process}.
For three applications, the analysis failed due to memory limitations when converting the APK file to Java bytecode by the \emph{dex2jar} library.
Within the remaining 97 applications, \SystemName~found a total of 1.135.441 string literals with non-zero length.
\SystemName~found at least one deobfuscation candidate for 79.272 string literals, indicating that these strings might be obfuscated.
Executing the program slices related to these 79.272 string literals resulted in 71.409 (90\%) of the executions in a string with non-zero length.
A manual review of the database content indicates a high success rate, however, we argue that it is infeasible to review all resulting strings.
Furthermore, some strings might falsely appear to be obfuscated (e.g., regular expressions), further complicating a manual review.

\begin{figure}[t]
	\centering
	\includegraphics[width=.7\linewidth]{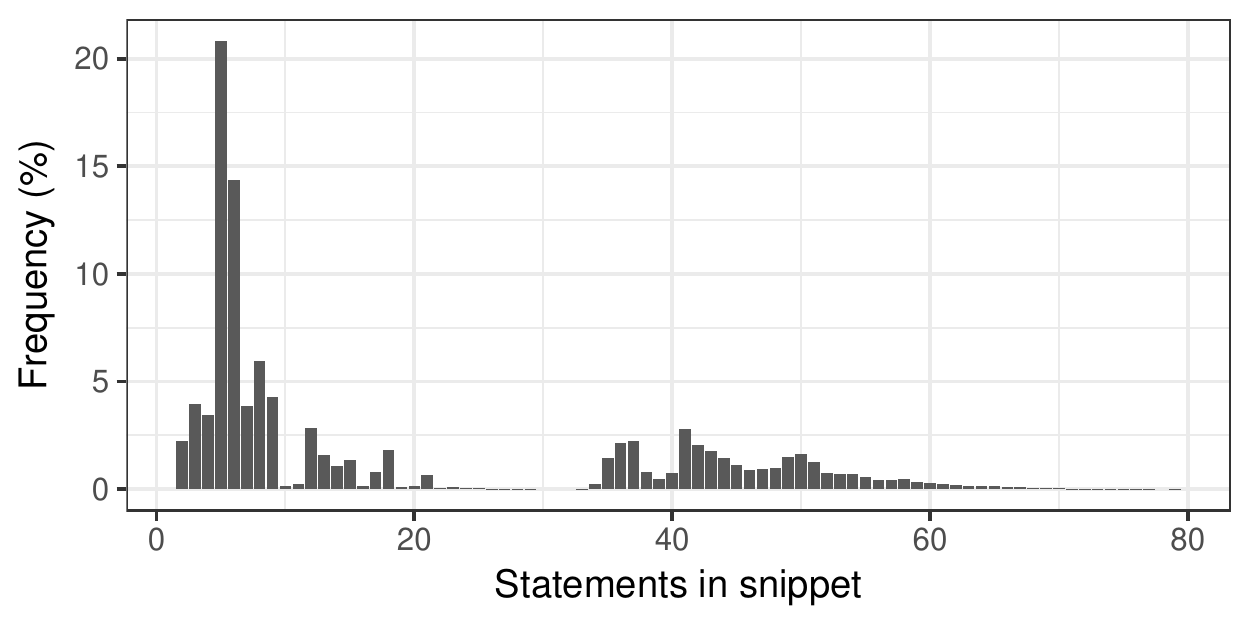}
	\caption{The distribution of the sizes of generated program slices. Only program slices which execution resulted in exit code 0 are included.}
	\label{fig:snippet_sizes_distribution}
\end{figure}

\begin{figure*}[t]
	\centering
	\includegraphics[width=\linewidth]{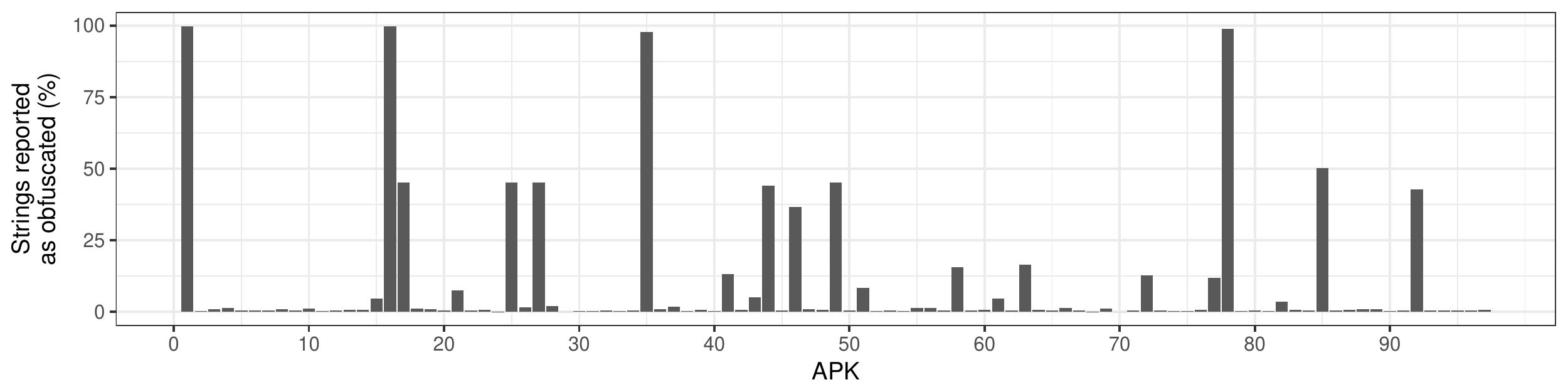}
	\caption{The distribution of ASCII characters in strings before deobfuscation (top plot) and after deobfuscation (bottom plot).}
	\label{fig:obfuscation_fraction}
\end{figure*}

\textbf{Character Distribution.}
To verify our tool, we consider the distribution of characters in the resulting strings that \SystemName~deobfuscated.
Our expectation is that the character distribution after deobfuscation resembles the frequencies of letters in the English language.
The distribution of ASCII character frequencies in the strings before and after deobfuscation is visualized in Figure~\ref{fig:character_distribution}.
The horizontal axis of Figure~\ref{fig:character_distribution} corresponds to the ASCII character index, ranging from 0 to 128, and the vertical axis shows the frequency of each character.
The plot only considers characters in the ASCII range and characters outside this range are ignored.
Note how the distribution of frequencies before deobfuscation resembles a uniform distribution.
Also observe that the frequency of characters in the range a-z have a higher variation, hinting at the presence of false positives (non-obfuscated string literals that \SystemName~reports as obfuscated).
The character frequency distribution after deobfuscation indicates a high frequency of characters in the range a-z, which is an indicator that our tool has successfully deobfuscated string literals to human-readable format.
An interesting observation is that there is a high frequency of ASCII character with index 32, which corresponds to a whitespace.

\textbf{Size of program slices}
Next, we evaluate the sizes of program slices generated by \SystemName.
We aim to distinguish the behaviour of different string obfuscators, based on the number of statements that a specific obfuscator inserts.
Figure~\ref{fig:snippet_sizes_distribution} shows the frequency of program slice sizes.
The plot only includes program slice which execution has resulted in a zero exit code.
The smallest program slice contains only two statements whereas the largest slice consists of 79 statements.
Note how over 20\% of all program slices contain exactly five statements.
Over half of all program slices, 59.0\%, contain ten or less statements.
Figure~\ref{fig:snippet_sizes_distribution} hints at the behaviour of different types of obfuscation, but requires a larger dataset and manual inspection to verify this hypothesis.

\textbf{Obfuscation usage.}
Finally, we highlight how the ratio reported obfuscated string literals behaves within the individual Android applications in our dataset.
Figure~\ref{fig:obfuscation_fraction} shows the percentage of string literals with at least one deobfuscation candidate, compared to the total number of string literals with non-zero length.
For each of the 97 APKs on the horizontal axis, we plot this percentage on the vertical axis.
Note how four APKs report percentages near 100\%, strongly suggesting that string obfuscation has been applied to all string literals in the source code.
We confirmed this by a manual inspection of these applications.
We also observe that many applications have not applied string encryption for the entire application, but only for the main application logic, excluding bundled third-party libraries and frameworks.
This decision is most likely motivated by performance reasons. 

\textbf{Implications.}
We argue that the resulting strings in non-obfuscated form provides crucial information to understand the application logic.
For example, a query for \enquote{pinning} amongst the deobfuscated strings quickly revealed the location of network logic in the source code of five different applications.
Similarly, a query for \enquote{root} and \enquote{check} reveals various checks for root permissions.
We believe that the resulting database can be a starting point for further (application-specific) analysis on security.

\textbf{Limitations.}
Although the aim of this work is to devise a generic mechanism for string deobfuscation, we identify three limitations of our approach.
First, \SystemName~fails when string obfuscation extracts the definition of the string literal from the method, and stores information on the string literal somewhere else.
Referring to Figure~\ref{lst:stringobfuscation}, an obfuscator might remove line 3, store the content of the encrypted string in a field, and apply reflection to retrieve the encrypted string.
\SystemName~would not be able to deobfuscate these strings since it cannot locate a string literal to deobfuscate.
Second, the proposed program slicing approach is optimistic and there is a probability that it fails to generate a correct, executable program slice when resolving undefined register values.
Finally, an obfuscator might intentionally insert code that requires an Android environment to prevent the deobfuscation from running on non-Android devices.
This limitation is easily countered by converting program slices to Dalvik bytecode and by conducting step \circled{7} in Figure~\ref{fig:deobfuscation_process} on an Android device.



\section{Related Work}
Cataclysed by the popularity of mobile applications, there have been several studies around the usage of code obfuscation techniques for both benign and malicious Android applications~\cite{Dong2018UnderstandingAO}~\cite{Wermke2018ALS}.
These studies show that string obfuscation is more commonly performed by malware developers, in order to avoid malware detection mechanisms.

Su et al., apply deep learning to deobfuscate Android applications and evaluate their approach on obfuscations performed by ProGuard and Allatori~\cite{Su2017DeobfuscatingAA}.
The work of Baumann et al., presents an approach for the automated deobfuscation of identifiers in Android applications that are renamed by the protection tool ProGuard~\cite{Baumann2017AntiProGuardTA}.
DeGuard, presented by Bichschel et al, use a statistical learning approach to reverse identifier obfuscation of commonly used data elements in Android applications.
DeGuard, however, is unable to reverse cryptographic obfuscation mechanisms, including string obfuscation~\cite{Bichsel2016StatisticalDO}.
The work of Yoo et al., relies on runtime analysis to extract deobfuscated strings and then traces these strings back to the point where the obfuscated string is defined~\cite{yoo2016string}.
In comparison, our approach is not dependant on the Dalvik runtime and does not require the analyst to run the Android application.

The popularity and fragmentation of the Android platform has resulted in much research around security analysis of mobile applications using reverse-engineering techniques~\cite{Enck2011ASO}.
Most of these tools rely on static analysis of information flows within an application~\cite{Qiu2018AnalyzingTA}.
The FlowDroid analyser assesses the security of mobile applications by using taint analysis, a popular method to determine which variables can be influenced by user input~\cite{Arzt2014FlowDroidPC}.
Amandroid, described by Wei et al., uses inter-component data flow analysis to reveal security issues and privacy violations~\cite{Wei2014AmandroidAP}.
Closely related to FlowDroid, DroidSafe combines a model of the Android runtime with static analysis design decisions to report on potential leaks of sensitive information~\cite{Gordon2015InformationFA}.

The technique of program slicing was introduced by Mark Weiser in 1981~\cite{Weiser1981slicing} and has widely been used to aid debugging, to perform data flow analysis, and to localize faults.
Specifically, our work has similarities with the work of Glanz et al., that also uses program slicing to show that string obfuscation is easily reversible~\cite{glanz2020hidden}.
A widely adopted, yet computationally expensive, approach to generate these slices is by constructing a program dependency graph (PDG) for each method in the application~\cite{Ottenstein1984ThePD}.
PDGs have proven to be useful when detecting code duplication, which is closely related to the approach in this work to identify the usage of string obfuscation~\cite{Krinke2001IdentifyingSC}.

Various tools exists, specifically targeted at Android applications, that adopt program slicing.
Hoffmann et al. introduces SAAF, a static analysis framework for Android applications, which creates program slices in order to perform data-flow analysis~\cite{Hoffmann2013SlicingDP}.
SAAF, however, does not consider slice generation in the presence of conditionals.
R-droid, created by Backes et al., employs slicing-based analysis to generate program slices that are easier to understand and validate~\cite{Backes2016RDroidLA}.
Zhang et al., present AppSealer, which uses program slicing to detect component hijacking and for automated patch generation~\cite{Zhang2014AppSealerAG}.
Azim et al., devised a dynamic program slicing model that handles asynchronous callbacks~\cite{Azim2019DynamicSF}.
Program slicing is pipelined in the AT2 tool used to analyse Android applications~\cite{Arshad2014AndroidAA}.
Feichtner applies static forward and backward slicing to identify misapplied cryptographic operations in Android and iOS applications~\cite{feichtner2019comparative}.
Although program slicing has been applied to mobile applications, this work is the first to apply slicing to deobfuscate strings, to the best knowledge of the authors.

\section{Conclusions}
We have presented \SystemName, a tool for deobfuscation of string literals in Android applications.
In contrast to existing tools, \SystemName~does not require prior knowledge on the type of obfuscation used.
\SystemName~uses program slicing to devise an executable code snippet with all relevant statements that contribute towards the deobfuscation of a specific string literal.
Since the cyclomatic complexity of implemented methods can be considerable, we adopt an opportunistic and lightweight program slicing algorithm, which can be considered as a hybrid solution between static and dynamic slicing.
By execution of the devised program slices, we obtain the string in non-obfuscated form.
The practicality of our work has been proven with an evaluation on 100 real-world financial Android applications.


\bibliographystyle{ACM-Reference-Format}
\bibliography{references}

\end{document}